# The air over there: exploring exoplanet atmospheres

Laura Schaefer[1], Vivien Parmentier[2]

1 Department of Geological Sciences, Stanford University, Stanford, CA 94305-2115

2 AOPP, Department of physics, University of Oxford, Oxford, UK

## Abstract

Atmospheric compositions for rocky exoplanets will depend strongly on the bulk planetary composition and the orbital position of the planet. Non-traditional gases may be present in the atmospheres of exceptionally hot planets. Atmospheres of more clement planets will depend on the abundances of volatiles acquired during planet formation and atmospheric removal processes, including escape, condensation, and reaction with the surface. While the observations of exoplanet atmospheres to date has focused on giant planets, a series of new space and ground-based observatories over the coming decade will revolutionize the precision and spectral resolution with which we are able to probe exoplanet atmospheres. This article consolidates lessons learned from the study of giant planet atmospheres, and points to the observations and challenges on the horizon for terrestrial planets.

## Introduction

The air we breathe on the Earth today is the product of billions of years of coupled planetary and biological evolution. The Earth's history can be divided into times before and after the Great Oxidation Event (~2.45 Ga), when $O_2$ first became abundant in the atmosphere. In the beginning, our atmosphere was different, probably dominated by $CO_2$ and $N_2$, as are Venus and Mars, though likely in different proportions. Over time, processes like escape, surface reactions, photochemistry, cloud formation, plate tectonics, and biology have changed the composition of our 'air'. These processes and many others occur on our neighboring planets and modify their 'air'; so too must these processes affect exoplanets.

Observations of rocky exoplanets are driven by questions such as: does planet A have plate tectonics? does planet B have life? Characterizing their atmospheres (e.g., composition, structure, and dynamics) gives us a single snapshot of their evolution; with sufficient understanding of the processes that shape atmospheres, we may be able to answer such questions. While the road to atmosphere characterization is challenging and interpretation will be difficult, the rewards of learning about the diversity of rocky planets will be great.

## Atmosphere Formation and Evolution

Planets form in a disk of gas and dust through both solid body collisions and gravitational attraction of gas. Dust condenses from the gas, so the temperature structure of the disk leads to compositional gradients in dust, with silicates and metals forming closer to the star and ices only forming further away. Exoplanet observations confirm that planets can migrate within the disk while forming, so volatile-rich planets may end up much closer to their star than they started out (Lin et al. 1996). Gravitational interactions in the disk can lead to scattering and highly eccentric orbits, which allows delivery of volatile materials formed further out to the inner planets. These processes complicate our interpretation of where and how an exoplanet formed based solely on its mass, radius, and inferred volatile budget.

Rocky planets as we know them in the Solar System have very limited volatile reservoirs compared to the gas and ice giants. The Earth has only 0.02 wt% water on the surface, although more may be stored in the mantle and core. Our atmosphere accounts for only ~1% of the total planetary radius, whereas some rocky exoplanets may have much more extended atmospheres. Volatile budgets of rocky exoplanets may span a larger range and much larger radius, but we focus on familiar processes relevant to Earth-like volatile budgets below.

### Atmosphere Formation

Atmosphere formation on rocky planets occurs throughout accretion as a result of many processes (see Figure 1), including (but not limited to): outgassing of the interior, impact degassing, gas accretion, and magma ocean ingassing. Other processes may remove atmosphere, including photoevaporation, impact erosion, hydrodynamic escape, and mass loss driven by self-luminosity following removal of the confining pressure of the protoplanetary nebula (often referred to as 'core-powered mass loss'). The net sum of these processes determines the thickness and composition of a planet's atmosphere.

Initial volatile accretion to objects probably begins with ices (e.g. $H_2O$, $CO_2$). As objects grow larger, heating from impacts, short-lived radioisotopes, and gravitational contraction melts the ices, which react with rocky/metallic materials to form oxidized or hydrated minerals. Small objects with minimal gravity lose volatiles via melting, sublimation and impact ejection. Planetesimals near each other can therefore have a diversity of volatile contents. Scattering during impacts and gravitational interactions drives further mixing across the disk. Larger objects gravitationally attract gas from the disk, making thin primary $H_2$ atmospheres. Objects forming in regions of high dust fraction grow rapidly and begin to accumulate gas at a faster rate than solids. These objects experience runaway growth and become gas giant planets, with thick primary $H_2$ atmospheres. Large planets often open gaps in the disk that isolate different regions and impede migration of volatile-rich pebbles to closer-in planets.

Rocky planets remain small and take longer (50 – 100 Myrs) to grow to Earth-masses, although Mars, which has been characterized as a stranded embryo, likely grew relatively quickly (<10 Myrs). It remains to be shown whether all rocky planets form in a similar manner to the Solar System planets. Thin primary atmospheres may be stripped from them as the disk gas disperses, around 3-5 Myr. Hot planets could ingas some disk gas into their molten interiors, but volatile isotopic compositions (e.g. $^{20}Ne/^{22}Ne$, $^{3}He/^{22}Ne$, $^{129}Xe/^{130}Xe$) in the Earth's mantle suggest that ingassing contributed <10% of Earth's volatile inventory (Marty 2012).

Impacts onto growing planets lead to partial devolatilization once the impact velocity exceeds critical values (~5 km/s for rocky materials, or 2-3 km/s for icy materials). During earlier stages, volatile material may be buried within the planet, to be released later during melting and outgassing. Impact degassing can produce a wide range of gas compositions depending on the composition of the impactor. Impacts can sometimes remove more gas than they deliver; erosion vs. retention for giant impacts depends on the presence of an ocean (Zahnle et al. 2007).

Giant impacts are the final stages of rocky planet growth. Collisions between similar sized objects lead to large scale melting. Magma oceans (i.e. large scale, partially-to-fully molten silicate regions) have different dynamical states depending on the scale, the degree of melting, which controls viscosity, and on the presence or absence of an atmosphere. Buried volatiles can be outgassed at this time into a massive atmosphere. The noble gas ratios of Earth's mantle suggests that Earth may have experienced multiple magma oceans (Tucker & Mukhopadhyay 2014). The radiative properties of the atmosphere determine when (or if) the magma ocean cools and solidifies (Zahnle et al. 2007), with common strong greenhouse molecules like water and $CO_2$ extending lifetimes. Strong early radiation and particle fluxes from young stars may drive atmospheric escape, which is more important for smaller planets on closer orbital periods, especially around stars smaller than the Sun, which remain very active for much longer.

After the giant impact stage, a long tail of late accretion is likely to occur in the Solar System. Impacts of material to a colder atmosphere may result in transient and localized atmospheric compositions that vary substantially from the existing background atmosphere (Zahnle et al. 2007). Within the Solar System, a much larger spike in late impacts (~3.9 Gyr), known as the Late Heavy Bombardment (LHB), was inferred from crater ages of the Moon. The LHB may have occurred due to a late instability in the orbits of the giant planets. The reality of the LHB has been called into question (Boehnke & Harrison 2016), but a late tail of primary accretion remains plausible. An event such as the LHB –the result of stochastic orbital interactions– should not be assumed for other planetary systems without good cause.

## Atmosphere Evolution

While a planet's initial atmosphere is set by the processes discussed above, it does not remain fixed. The Earth's atmosphere has gone through major changes in composition and possibly thickness over its history as a result of many internal and external processes. External influences include stellar radiation and particle fluxes (e.g. solar wind), impacts, and energetic particles from neighboring stars. Internal influences include continued release of gravitational energy, internal convection, and material exchange between envelope and interior layers, true for both giant and rocky planets.

Atmospheric escape has removed much of their initial water from both Venus and Mars, and left an imprint on the Earth's atmosphere as well. Atmospheric escape by photoevaporation or core-powered mass loss may be responsible for sculpting the observed small exoplanet demographics into a bimodal distribution of planets with massive atmospheres (1.5 – 3 $R_{Earth}$), which cannot be definitively identified as rocky planets, or planets with relatively thin atmospheres (< 1.5 $R_{Earth}$), which are recognizable as rocky planets plausibly similar to those in the Solar System (Fulton et al. 2017).

At the end of a magma ocean phase on the Earth, models predict the collapse of a steam-dominated atmosphere to form the earliest ocean. A massive residual atmosphere of $CO_2$ would have dissolved in this early ocean and reacted with hot young crustal rocks to form carbonates, which could have efficiently removed 50-200 bars of $CO_2$ down to 1-2 bars in the remaining atmosphere (Zahnle et al. 2007). The carbonate-silicate weathering cycle continues and may be a thermostat for planets with stable surface water due to the temperature-dependence of the weathering reaction.

Other volatiles are also removed by reaction with crustal rocks. Water in the ocean can hydrate silicates (serpentinization). Tectonics may lead to deep burial of volatiles or even transport into the mantle through subduction, which will govern the long-term availability of volatiles such as water, $CO_2$ and $N_2$ (Sleep et al. 2012).

Stellar radiation drives photochemistry in the middle and upper atmospheres. On Earth today, photochemistry produces the ozone layer, which absorbs most of the Sun's UV radiation, and creates a temperature inversion in the stratosphere. Photochemistry in Venus's atmosphere produces sulfuric acid clouds. On Earth, condensation clouds of water form in the lower atmosphere due to temperature decreasing with height. Other condensable components, including various acids (sulfuric, nitric, and hydrochloric) and organic aerosols, also form under different conditions in Earth's atmosphere. $CO_2$ and $H_2O$ condensation clouds form on Mars, and many other exotic condensation clouds are predicted in both giant and rocky exoplanet atmospheres in different temperature regimes (Mbarek & Kempton 2016).

## Rocky Planets: From Hot to Cold

Rocky exoplanets occupy orbital, radiation, and compositional spaces previously unexplored (see Fig. 2). However, lessons learned from familiar planets help us classify, model, and understand these foreign worlds. Below, we discuss some of these new types of worlds.

### Lava Worlds: Boiling hot

The first rocky exoplanets discovered were two similar planets: CoRoT-7b and Kepler-10b. Both planets orbit Sun-like stars with periods of only ~20 hours. These planets are very hot. The planets are likely **tidally-locked**, with dayside temperatures of 1800 – 2200 K, more than hot enough to melt rocks.

If these planets formed where we see them today, it's unlikely they would have accreted many (or any) volatiles. Even if they did, the close proximity to the star means that they are constantly being blasted by stellar wind particles and radiation, enough to strip initial volatiles. But that doesn't mean these planets have no atmosphere.

At these hot temperatures, new materials become 'volatile'. Dayside temperatures are hot enough to melt a pool of lava, thickest at the substellar point and shallower towards the limbs due to a strong temperature gradient (Léger et al. 2011). Vaporization of the lava produces an atmosphere with gradients both in pressure and composition, driven by fractional vaporization of the silicate melt at the surface. Dominant gases include Na, SiO, O and $O_2$. Pressure gradients lead to winds, which may whisk silicate gases around to the cold nightside of the planet where they would condense. Fractional vaporization and atmospheric escape could both gradually deplete the lava pool of more volatile components like Na, although it has yet to be shown if escape of silicate atmospheres is feasible.

### Searing-hot Planets

Most rocky exoplanets detected to date occupy orbital and stellar insolation spaces akin to Mercury. These planets have equilibrium temperatures low enough that stellar radiation alone won't melt the surface, but hot enough to sear a steak or maybe even melt lead! These planets are interior to the defined habitable zone (HZ); the inner edge is defined by the runaway greenhouse threshold, which means if these planets have any water, it would only exist in the atmosphere as steam. Steam can be split into H and O atoms by photolysis, and H will tend to escape, especially for hot rocky planets. These planets may therefore lose some or all of their water during their evolution (Luger & Barnes 2015). Massive outflows of H may also drag other gases along that otherwise might not escape, like O. This may be true for GJ 1132 b; models suggest that an early steam atmosphere could have left behind only relatively thin $O_2$ atmosphere (Schaefer et al. 2016). Ground based observations have so far not been able to confirm the presence of an atmosphere (Diamond-Lowe et al. 2018).

Like the boiling-hot planets, these planets may be tidally-locked. Like Mercury, a planet may be tidally-locked without being tidally-synchronized; Mercury is locked in a 3:2 spin-orbit resonance, rather than a 1:1 resonance, so tidally-locked exoplanets may not necessarily be synchronously rotating. For synchronously rotating planets, if they have an atmosphere, it transports heat between day and night, evening out temperature differences, similar to Venus. If they retain water, the greenhouse effect of a steam atmosphere would easily lead to a globally molten surface. Silicate materials dissolve readily in steam atmospheres, so gases like $Si(OH)_4$ may be present. Condensation clouds in these planets atmospheres may include species such as KCl and ZnS (Mbarek & Kempton 2016). An atmosphere composed only of $CO_2$ is likely insufficient to melt the surface.

Planets with little or no atmosphere will have huge temperature contrasts between day and night side. Observations of LHS 3844 b with the Spitzer Space Telescope confirmed that this planet has little if any atmosphere and has a surface that resembles basalt (Kreidberg et al. 2019). These observations cannot currently rule out an atmosphere as thick as Mars's, so follow-up observations will further test this model.

If water has been lost then, we may expect their atmospheres to be dominated by $CO_2$ and $N_2$ (Forget & Lecuyer 2014). These gases are heavier and produce larger fragments when photolyzed, so that escape of these components is likely less efficient. In a multicomponent atmosphere, water loss may leave behind residual $O_2$ gas.

## Exo-Earth or Exo-Venus: Just right?

Within the HZ, a diversity of rocky planets still exist, which may or may not have water. HZs move as the host star evolves; for M dwarfs, they move inwards over time, so planets in the HZ today had a hot, uninhabitable start. Planets that start with water but lose hydrogen may have massive abiotic $O_2$ atmospheres (Luger & Barnes 2015), which are potential biosignature false positives. However, reaction of $O_2$ atmospheres with FeO in silicates (Schaefer et al. 2016) may help reduce this false positive, which may partially explain the lack of $O_2$ on Venus.

The composition of the initial secondary atmospheres for rocky planets will depend on relative abundances of the volatile elements (H/C/N/S) and the oxidation state of the planet's interior, which controls volcanic gas abundances. Oxidized planets will outgas $H_2O$, $CO_2$ and $N_2$, whereas reduced planets will outgas mostly $H_2$, CO, $CH_4$, and $NH_3$. Degree of outgassing may be suppressed on more reduced planets by formation of refractory phases, such as carbides, sulfides and nitrides. At habitable temperatures, water should condense and make oceans. Even reduced planets could have liquid water at their surfaces. $H_2$ is an efficient greenhouse gas that may even extend the HZ (Pierrehumbert & Gaidos 2011) beyond its traditional boundaries.

Life on Earth modifies the Earth's surface and atmosphere in a multitude of ways, including production of $O_2$ and $CH_4$, which are in disequilibrium in the Earth's atmosphere. Detecting life on other planets means detecting molecules produced by biological reactions in the atmospheres of exoplanets. Multiple gases out of equilibrium with each other (Krissansen-Totton et al. 2016) is perhaps the best signature of life, but one that requires us to understand the processes operating in these atmospheres and eliminating all other possible means of production to be certain of their origin. Better understanding of the processes involved will be needed before confident detection of a biosignature can be made. More discussion of biosignatures is given in Rimmer et al. (this issue).

The Trappist-1 system, with seven planets ranging from very hot (b & c) to very cold (f & g), represents our best chance of testing models of rocky exoplanet atmospheres. New analysis of mass and radius measurements suggests that all of the Trappist-1 planets are consistent with a core mass fraction (CMF) of only 17% (compared to 33% for the Earth) and minimal volatiles (Agol et al. 2020). Degeneracies inherent in using mass-radius data to determine internal structure means that the Trappist-1 planets could either be uniformly depleted in Fe relative to the Earth, or alternatively an otherwise Earth-like CMF with higher volatile fraction than the Earth. Trappist-1e is the most Earth-like planet in this system, with a mass slightly lower than Venus and an equilibrium temperature very close to the Earth's. None of these planets likely retain a primary $H_2$ atmosphere, but higher mean molecular weight atmospheres composed of gases like $H_2O$, $CO_2$, or $N_2$ are plausible (Turbet et al 2020). Measurements with upcoming space telescopes will give us our first insights into the atmospheres of HZ planets.

## Exo-Titan and Beyond: Cold

Only a few known small exoplanets have been found beyond the HZ, like Trappist-1g and f. Objects like Titan and Pluto tell us that small cold planets can maintain sizable atmospheres, but they are very unlike more traditional rocky planets. At these distances, water is no longer a volatile species, behaving more like silicate rocks for HZ planets. Planet formation models suggest that objects formed outside of the snow line will have equal (or greater) amounts of ice to silicate and metal. Planets the size of Earth or larger formed beyond 5 AU (or closer, given sufficiently low early XUV radiation) may be able to hold on to their primary atmospheres of $H_2$, which might provide some warming of the surfaces. Other gases like $N_2$, $NH_3$, and $CO_2$ become less volatile at greater distances and may be stable as ices or liquids at planetary surfaces. These species that are at or near their sublimation/evaporation points now serve the same role in surface-atmosphere processes as water does on the Earth.

Considering how little stellar light they receive, photochemistry plays a surprisingly important role in the atmospheres of Titan and Pluto, producing organic hazes in their atmospheres. Observations of Titan with Cassini designed to mimic exoplanet transit spectroscopy were able to retrieve wavelength-dependent properties of the haze (Robinson et al. 2014).

## Observing Exoplanet Atmospheres

### How do we observe exoplanets atmospheres?

Observing exoplanet atmospheres is hard. Planets are much closer to their stars than their stars are to us. To see a planet's atmosphere, we first need to remove the star. This has been done in three main ways: temporal, spectral or spatial separation (see Fig. 3).

**Temporal separation**

Temporal separation is very efficient for planets with short periods. Planets in wide orbits have a much smaller transit probability (e.g. < 1% for an Earth-like orbit), and the rarity and length of their transits impose large constraints on telescope scheduling (Earth transits once a year for 12 hours). When a planet passes behind its star (a secondary eclipse), we can measure the spectrum of the star alone and subtract it from the combined stellar and planetary flux observed at other times. When the planet transits the star, we can observe how the planetary atmosphere filters the stellar light and obtain a transmission spectra. Transit spectroscopy can be done successfully with small telescopes and can be performed from space (currently the largest space telescope is the 2.5m Hubble Space Telescope). However, subtraction of the stellar signal from the planetary signal is often complex and can lead to significant uncertainties and false positives.

**Spectral separation**

Spectral separation uses the velocity difference between the star and the planet. At high spectral resolution (R~100,000), spectral lines from the planet and stellar atmospheres are Doppler-shifted differently (see Fig. 3), allowing one to recover the planet's spectra. Many photons and an extremely stable instrument, with precisely controlled temperature (<0.001K), are needed to obtain such high spectral resolution, so this technique is confined to ground based telescopes, which brings all the difficulties of removing the Earth's atmospheric contamination.

**Spatial separation**

Spatial separation means taking a direct image of the planetary system. The telescope has to both spatially separate the star and planet and also handle the large brightness contrast between the two. Up to now, direct imaging has only been applied to young (hence self-luminous) giant planets in orbits wider than Jupiter's. Because spatial resolution is inversely proportional to telescope diameter, a very large telescope is required, so direct imaging has only been used from ground based observatories so far.

## From observations to abundances

Numerous exoplanet atmospheric spectra have been detected by the space-based Hubble and Spitzer observatories (for transiting planets), and the largest ground based telescopes (for spectral and spatial separation techniques). Two species found in almost all exoplanet atmospheres are water and aerosols (Kreidberg et al. 2018). High resolution spectroscopy allowed the detection of additional gases, such as $CH_4$, CO, TiO, He, H, Na, K and Fe (Birkby et al. 2018).

To measure absolute abundances, particularly for minor constituents (e.g. $CO_2$ in Earth atmosphere), one must know what atmospheric depths the observations are probing. Measurements of the shape of observed spectral features can tell us which pressure levels are probed. This requires a much higher precision than 'simple' species detection, which uses the size of the spectral features. Numerous physical mechanisms, from instrumental systematics to planetary complexity, can alter the shape of the spectrum, which complicates abundance determinations. Obtaining molecular abundance ratios is theoretically simpler, since it relies on the relative sizes of spectral features and averages out the effects that alter the shape of molecular features.

One approach used to solve the inverse problem is an atmospheric retrieval. Thousands to millions of forward models are run, and complex statistical techniques use these models to constrain the true atmospheric properties. Although the method provides a quantitative estimate of chemical species abundances, it has its limitations. A real planet is much more complex than can ever be fully modelled. We therefore simplify the problem and choose which parts of the planet model to include and which to neglect. All published conclusions, particularly measured mixing ratios, are valid only if none of the model assumptions affect the result.

We have so far obtained spectral information from planets ranging from temperate Neptune-like worlds to Jovian worlds hotter than some stars. We have learned that temperature plays a fundamental role in setting the atmospheric properties of these objects. Ultra hot Jupiters ($T_{eq} > 2000K$) are so hot that water thermally decomposes on their dayside. Hot Jupiters ($T_{eq} = 1000 - 2000K$) have partially cloudy daysides, cloudy nightsides and variable cloud composition with temperature. Most have a composition compatible with their host star composition, albeit with large uncertainties. Cooler planets for which we have spectral information are often less massive and orbit cooler stars. This population appears more cloudy or hazy, making accurate determination of their abundances even more challenging. It is therefore complex to tease out if the differences we see are due to the variation of planet temperature, planet mass or stellar type.

Current atmospheric measurements therefore cannot reliably answer planet formation questions. This will likely change in the coming decade. The James Webb Space Telescope will provide observations 100 times more precise for 10 times more planets. JWST will have the spectral coverage and precision needed to move from simple detection to quantification of molecular abundances. Additionally, ground based observations at high resolution will allow more atomic and molecular species to be detected. The relative abundances of rocky material (such as silicate or magnesium), metals (such as iron), alkalis (such as sodium and potassium), and ices (such as oxygen, carbon, nitrogen) in (ultra-) hot Jupiters will inform us much more precisely about the formation processes of these planets.

## Future Challenges for Rocky Exoplanets

Over the past 20 years of giant exoplanet observations, we have learned that it is easy to have precise but incorrect conclusions. Although atmospheric retrievals report molecular abundances to within one order of magnitude, they can differ by several orders of magnitude when different assumptions are made. Planetary atmospheres are complex, 3D objects, and the size of the signals we are looking for, particularly for rocky planets, will often be comparable to the various noise sources. Stellar jitter, for instance, imparts radial velocity noise on the order of the radial velocity signal of Earth orbiting the Sun. Errors include both false positives, when a molecule is detected but not truly present, and biased abundances, when a molecule is truly detected but the inferred abundances are different from the true value. We list below a few mechanisms that may trouble future rocky planet retrievals. These are limited by our imagination and experience, and we have no doubts that new observations will bring new ways of being wrong.

- Inhomogeneities in the stellar surface (e.g., star spots) can induce spurious molecule detections in transit spectra. A species present in the stellar photosphere (e.g. water in M dwarfs, sodium or metal oxides in G stars) can easily be imprinted in the transmission spectra (Rackham et al. 2018).
- Mass and radius errors can cause misinterpretations. For transmission spectra, uncertainty in the planet's gravity produces a similar uncertainty in the mean molecular weight of the atmosphere, reducing our ability to detect if a spectrally inactive background gas (e.g., $N_2$) is present (Batalha et al. 2019).
- An undetected moon, whose spectrum combines with the planetary spectra, can mimic disequilibrium chemistry in an atmosphere (Rein et al. 2014).
- In transmission, emission, or reflected light observations, aerosols hide the deeper atmosphere. Only the tops of the clouds of Venus can be seen, except at specific microwave wavelengths, leading many to think that Venus had a habitable environment for decades. When the aerosol cover is patchy, it can mimic the presence of a higher mean molecular weight, affecting all measured molecular abundances (Line and Parmentier 2016).
- Instrumental and astrophysical systematics can change the shape of the spectrum. While it is tempting to observe enough times to detect a spectral feature, abundance measurements depend on the shape of the spectral feature, which requires a higher precision, and is not improved by excessive observations. To date many molecular inferences are based on retrievals that do not correctly fit the data (e.g. that have very low p-values), which could be due to unaccounted instrumental noise or an undetermined astrophysical mechanism. Both issues can encourage modelers to add exotic species to mimic the noise in the spectra.
- Horizontal and vertical variations in abundances, plausibly due to photochemistry (e.g. the ozone layer on Earth) or large thermal contrasts, can significantly alter the shape of the spectra. When interpreted with a 1D framework assuming that abundances are constant with altitude, this can lead to orders of magnitude errors in the retrieved abundances of hot Jupiters (Parmentier et al. 2018).

- Time variability might be a strong issue. Most proposals to obtain a significant spectrum of a rocky planet, particularly for temperate ones, involve summing up tens of transits over several years or integrating for weeks or even months for a directly-imaged planet to obtain the necessary signal to noise ratio. If a planet spectrum varies significantly during that period, e.g. Mars during a dust storm, then the observed average spectrum will not be well represented by a 1D model.

The coming generation of telescopes will give us the best opportunity to understand these planetary atmospheres. Emission spectra of very hot planets, accessible by the MIRI instrument on the James Webb Space Telescope, will be modulated by absorption by their tenuous atmosphere (with signatures of e.g. SiO, suggested by Léger et al. 2011) and spectral variation of different types of rocks on their surface (tentative signatures of e.g. granitic or basaltic rocks, suggested by Kreidberg et al. 2019). For cooler planets, transmission spectra with JWST could detect the presence of water, $CH_4$ or $CO_2$ and the size of these features could constrain the mean molecular weight of the atmosphere (if cloud-free) (Morley et al. 2017).

Further in the future, large ground-based telescopes such as the E-ELT (under construction) or the space-based LUVOIR telescope (under selection) could lead to the first detection of biosignature gases such as $O_2$ (see Komacek et al. in this issue for more details).

Observations of rocky exoplanets will allow us to test a variety of formation and evolution models. Confirming whether lava worlds are completely volatile depleted would help discriminate between models of in situ formation or migration. Measuring surviving water concentrations on searing hot planets would help test models of atmospheric escape. Finding trends in atmospheric oxidation state with orbital distance, stellar type, or planet mass would test theories regarding atmospheric escape, as well as oxidation gradients within a protoplanetary disk.

For temperate planets, the concentration of $CO_2$ could help us determine whether liquid water is present at the surface, since $CO_2$ should be kept to very low abundances in the atmosphere of a rocky planet undergoing the silicate weathering cycle, as it is on the Earth. Building a statistically significant sample of planets on which such measurements have been made will immeasurably improve our understanding of planet formation and evolution.

The rocky planets that will be observed in the coming decade (see Fig.2 for planned JWST observations) span a diverse range of temperature, masses, radii, and stellar type. Each of these worlds will be unique. No single set of assumptions could correctly interpret the spectra of such a diverse set of planets. All molecular detections and all retrieved abundances will be subject to complex biases that must be carefully evaluated before trusting any emerging trend or drawing firm conclusions. The characterization of giant exoplanets is leading the way. JWST and later the ARIEL telescope will provide a much more detailed view of hot and warm atmospheres, allowing us to test how far off our current inferences of these atmospheres are. The lessons learned will be crucial to correctly interpret the much noisier spectra of rocky planet atmospheres.

## References (limit 30):

# Figure Captions

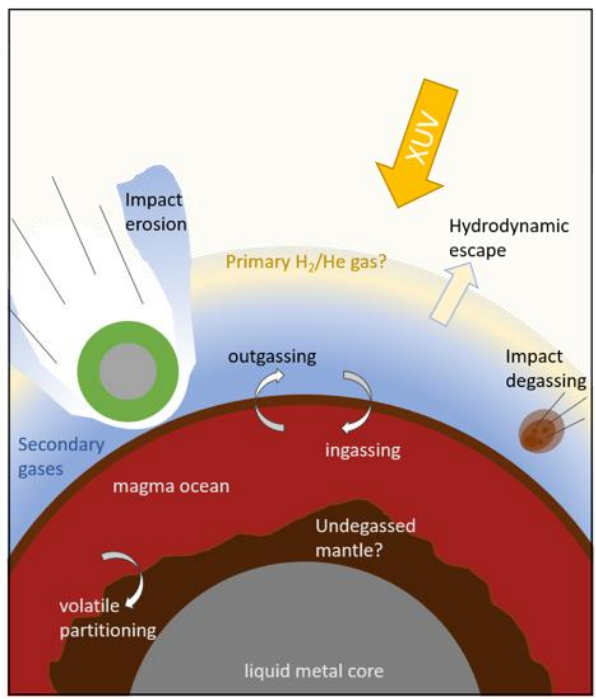

**Figure 1:** Atmospheric formation processes include accretion of a primary nebular gas envelope, impact degassing of infalling planetesimals, and outgassing of volatiles buried in the planet's interior. Magma oceans facilitate outgassing and also allow ingassing of some nebular gases. Volatile partitioning between solid mantle and magma ocean, between magma ocean and atmosphere, and between mantle and core all influence the distribution of volatile elements. Hydrodynamic escape of the massive early atmosphere can be driven by early large XUV fluxes from the host star.

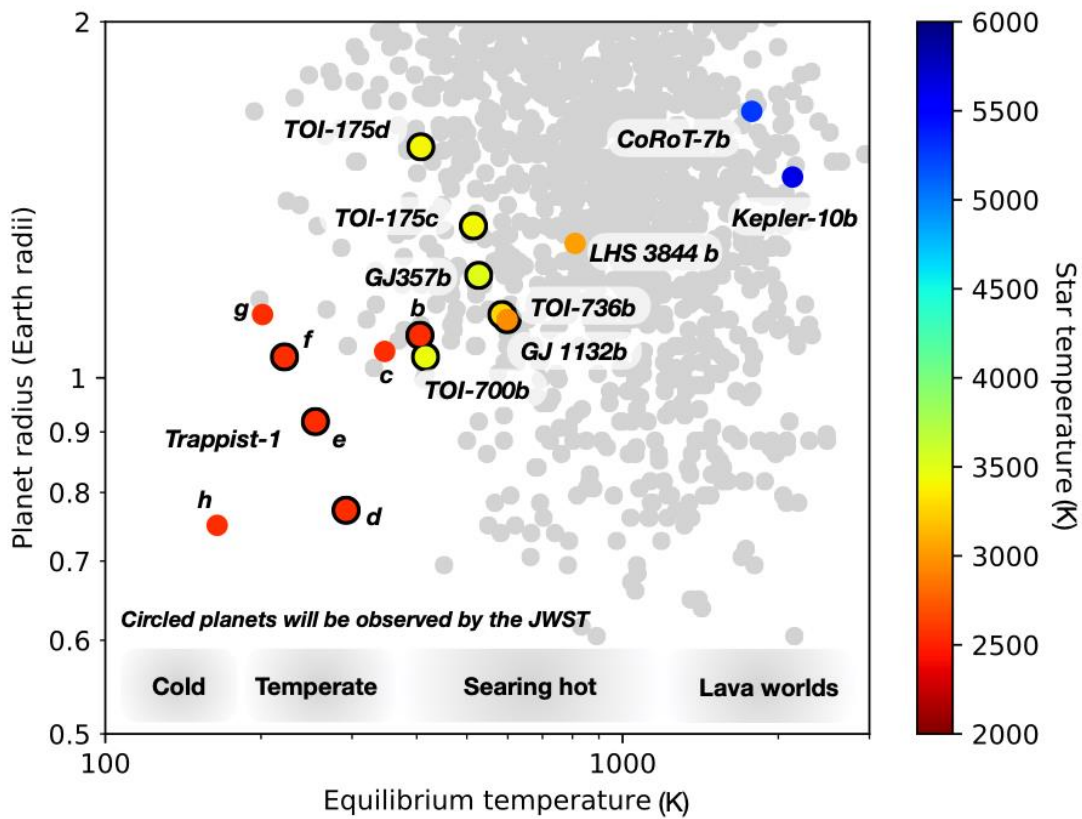

**Figure 2:** Known rocky planets' radii as a function of their equilibrium temperature calculated assuming zero albedo. Planets discussed in this article are color-coded by their stellar temperature. Planets circled in black have atmospheric observations planned by the James Webb Space Telescope.

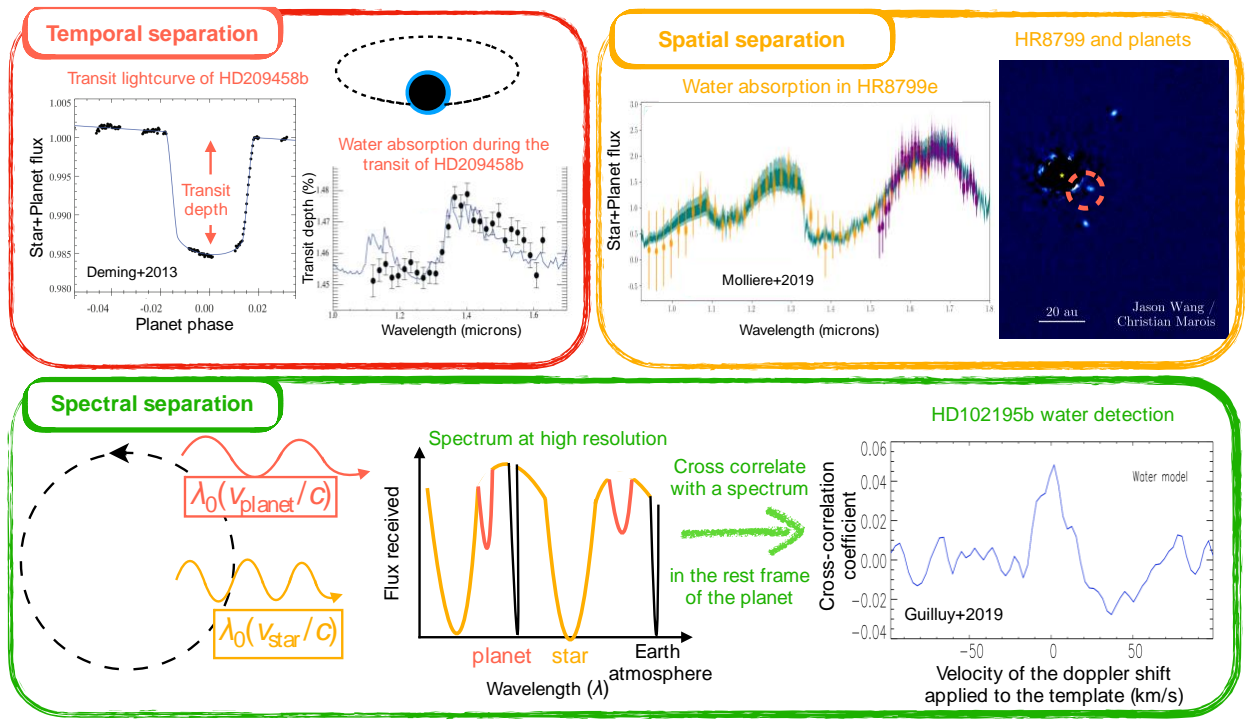

**Figure 3:** Summary of the different techniques used to observe exoplanet atmospheres. *Left* Temporal separation allows us to obtain the transmission spectrum of the planet by looking at the wavelength-dependent transit depth. *Top Right* Direct imaging can be used to obtain spectra of planets in wide orbits. *Bottom* The spectroscopic separation technique. Planet and star have different velocities with respect to Earth, leading to separation of their spectral lines by different Doppler shifts. Given that the signal-to-noise ratio is often too small to resolve individual lines, the spectrum is usually cross-correlated with a template of spectral lines shifted in wavelength. If the maximum of the cross-correlation coefficient happens at a shift corresponding to the Doppler shift expected from the exoplanet orbit, molecules are detected.